\documentclass[aps,prl,reprint,superscriptaddress,noeprint]{revtex4-2}

\usepackage[english]{babel}

\usepackage{amsmath}
\usepackage{amssymb}
\usepackage{wasysym}
\usepackage{graphicx}
\usepackage{hyperref}
\usepackage{xcolor}
\usepackage{physics}
\usepackage{units}
\usepackage[normalem]{ulem}

\usepackage[capitalise]{cleveref}  % /cref{fig1} will print 'Fig. 1'

\begin{document}

\title{Black-body radiation induced facilitated excitation of Rydberg atoms in optical tweezers}

\author{Lorenzo Festa}
\affiliation{Max-Planck-Institut für Quantenoptik, 85748 Garching, Germany}
\affiliation{Munich Center for Quantum Science and Technology (MCQST), 80799 München,  Germany}

\author{Nikolaus Lorenz}
\affiliation{Max-Planck-Institut für Quantenoptik, 85748 Garching, Germany}
\affiliation{Munich Center for Quantum Science and Technology (MCQST), 80799 München,  Germany}

\author{Lea-Marina Steinert}
\affiliation{Max-Planck-Institut für Quantenoptik, 85748 Garching, Germany}
\affiliation{Munich Center for Quantum Science and Technology (MCQST), 80799 München,  Germany}
\affiliation{Physikalisches Institut, Eberhard Karls Universität Tübingen,  72076 Tübingen, Germany}

\author{Zaijun Chen}
\affiliation{Max-Planck-Institut für Quantenoptik, 85748 Garching, Germany}
\affiliation{Munich Center for Quantum Science and Technology (MCQST), 80799 München,  Germany}

\author{Philip Osterholz}
\affiliation{Max-Planck-Institut für Quantenoptik, 85748 Garching, Germany}
\affiliation{Munich Center for Quantum Science and Technology (MCQST), 80799 München,  Germany}
\affiliation{Physikalisches Institut, Eberhard Karls Universität Tübingen,  72076 Tübingen, Germany}

\author{Robin Eberhard}
\affiliation{Max-Planck-Institut für Quantenoptik, 85748 Garching, Germany}
\affiliation{Munich Center for Quantum Science and Technology (MCQST), 80799 München,  Germany}
\affiliation{Physikalisches Institut, Eberhard Karls Universität Tübingen,  72076 Tübingen, Germany}

\author{Christian Gross}
\email[]{christian.gross@uni-tuebingen.de}
\affiliation{Max-Planck-Institut für Quantenoptik, 85748 Garching, Germany}
\affiliation{Munich Center for Quantum Science and Technology (MCQST), 80799 München,  Germany}
\affiliation{Physikalisches Institut, Eberhard Karls Universität Tübingen,  72076 Tübingen, Germany}

\date{\today}

\begin{abstract}
%600 characters max.
Black-body radiation, omnipresent at room temperature, couples nearby Rydberg states.
The resulting state mixture features strong dipolar interactions, which may induce dephasing in a Rydberg many-body system.
Here we report on a single atom resolved study of this state contamination and the emerging pairwise interactions in optical tweezers.
For near-resonant laser detuning we observe characteristic correlations with a length scale set by the dipolar interaction.
Our study reveals the microscopic origin of avalanche excitation observed in previous experiments.
\end{abstract}

\maketitle

Tailored many-body systems can be engineered from atomic ensembles laser-coupled to Rydberg states.
In combination with optical tweezers this forms a versatile platform for quantum simulation and computation~\cite{saffman2010, browaeys2020, graham2019, levine2019, madjarov2020, ebadi2021, scholl2021}.
Understanding decoherence channels is of prime importance for these emerging  applications of Rydberg atoms.
At room temperature, black-body radiation is known to incoherently drive transitions between nearby Rydberg states, a process often dominating the decay rate of a Rydberg state~\cite{beterov2009}.
In a many-body setting, the resulting state contamination with Rydberg states of opposite parity opens extremely strong dipolar interaction channels.
These uncontrolled interactions lead to dephasing, which may severely limit the coherence time for ensembles of Rydberg atoms.
Previous works have observed and studied the presence of interaction induced dephasing and line broadening spectroscopically in a bulk setting~\cite{goldschmidt2016, desalvo2016, aman2016, gaul2016, boulier2017, dehond2020}. 
A scaling analysis~\cite{goldschmidt2016} and dynamic experiments~\cite{boulier2017} pointed towards black-body radiation induced state contamination to trigger an avalanche excitation process. 
Dipolar interactions cause level shifts, such that the normally off-resonant laser becomes resonant. 
This facilitated excitation results in quick population build up in the Rydberg state. 
Due to the high probability to undergo a black-body radiation induced state change, even more contaminant atoms are produced speeding up the facilitation process. 
Mean-field models have been employed to explain this effect, but they have shown large quantitative deviation from the data. 
This triggered a refined theoretical analysis pointing out the importance of correlations between the excited Rydberg atoms~\cite{young2018}. \\

Here we report on a study of the state contamination induced interactions with neutral atoms individually trapped in a two-dimensional optical tweezer array.
Similar to prior experiments in bulk, we near-resonantly laser-couple the atoms to a Rydberg state~\cite{goldschmidt2016, desalvo2016, aman2016, gaul2016, boulier2017, dehond2020}.
This realizes the setting of Rydberg dressing~\cite{santos2000, bouchoule2002a, henkel2010, pupillo2010, johnson2010}, a versatile strategy to realize complex Hamiltonians for the study of quantum magnets~\cite{vanbijnen2015, glaetzle2015,potirniche2017} and to prepare resource states for quantum metrology~\cite{bouchoule2002a, gil2014, kaubruegger2019, kaubruegger2021}.
Coherent evolution under Rydberg-induced interactions has been reported in small systems~\cite{jau2016,zeiher2016, zeiher2017,guardadosanchez2021} or for relatively short times~\cite{borish2020} and avalanche excitation has been observed as one limiting process~\cite{zeiher2016, zeiher2017}.
The single atom resolved tweezer system allows us to probe the excited Rydberg atoms one-by-one and to study the facilitation process in the pairwise limit outside of the avalanche regime.
We extract the characteristic correlation length and directly show that this matches the length scale set by the dipolar interactions.

\begin{figure}[t!!!]
\includegraphics[width=\columnwidth]{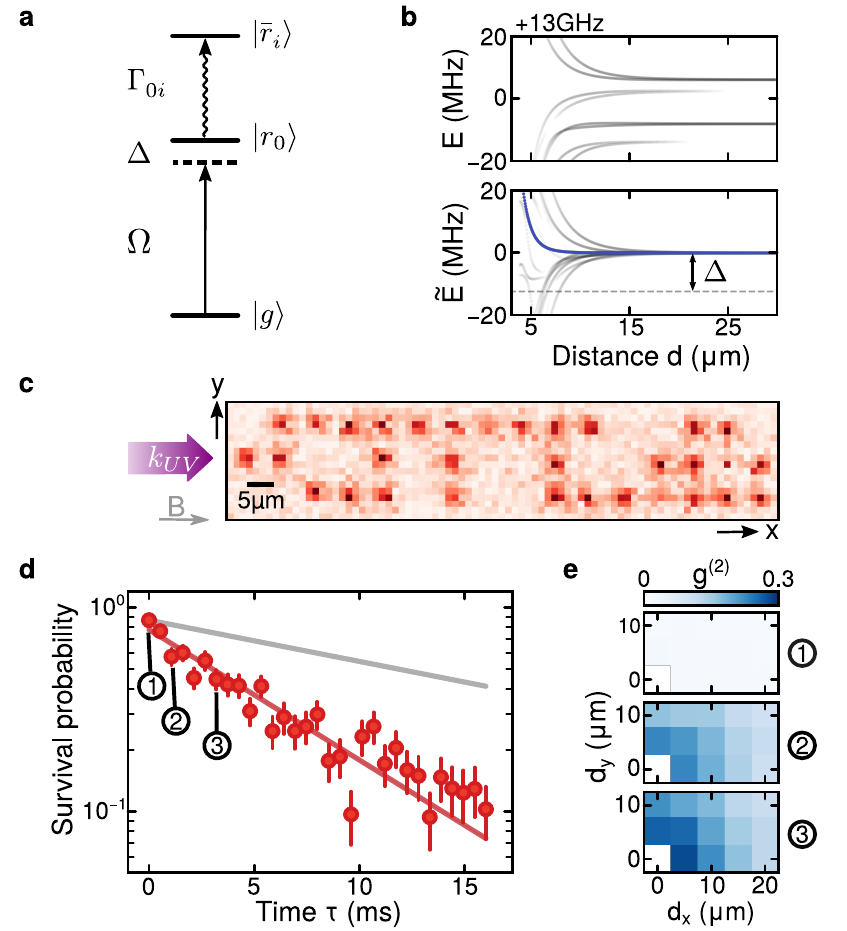}
\caption{
\textbf{a.} Simplified level scheme with the atomic ground state $\ket{g}$ and
laser coupled Rydberg state $\ket{r_0} = \ket{62P_{1/2}, m_J=-1/2}$ with Rabi
frequency $\Omega$ and detuning $\Delta$. Black-body radiation couples to
nearby Rydberg states $\ket{\bar{r}_{i}}$ with rate $\Gamma_{0i}$. \textbf{b.}
Van der Waals and dipolar pair-potentials. The dipole-dipole potentials between
$\ket{r_0}$ and magnetic substates of the $\unit[13]{GHz}$ higher lying,
strongest coupled states $\ket{61D_{3/2}}$ and $\ket{61D_{5/2}}$ are shown in
gray. The shading is proportional to the laser coupling rate for a magnetic
field of $\unit[10]{G}$, which mixes the finestructure states. In a black-body
event, the energy difference between $\ket{r_0}$ and $\ket{\bar{r}_{i}}$ is
provided by the microwave photon, effectively collapsing all pair-potentials
asymptotically (lower panel). The van der Waals potential between a pair of
atoms in the state $\ket{r_0}$ is highlighted in blue in the lower panel,
illustrating its much shorter range. \textbf{c.} Single fluorescence image of
ground state atoms in a 3$\times$16 tweezer array with $\unit[5]{\mu m}$
spacing. The magnetic field direction is indicated by the gray and the UV laser
direction by the purple arrow. \textbf{d.} Measurement of the trap lifetime in
the 3$\times$16 array at $\Delta=-2\pi \cdot \unit[4]{MHz}$ and $\Omega=2\pi
\cdot \unit[430]{kHz}$. The dark-red line shows an exponential fit to the data
revealing a trap lifetime of $\unit[6.8(4)]{ms}$, much shorter than the laser
phase noise limited trap lifetime of $\unit[21.4(1.3)]{ms}$ measured for
individual atoms (gray line)~\cite{si}. Error bars denote 1 s.e.m. \textbf{e.}
Two body correlator $g^{\mathrm{(2)}}$ at different illumination times marked with the numbers in d, showing the growth of correlations over time.
}
\end{figure}

In our experiment we use Potassium-39 atoms initially prepared in the $\ket{g} = \ket{4S_{1/2}, F=2, m_F=2}$ ground state.
We laser-couple the atoms to the $\ket{r_0} = \ket{62P_{1/2}, m_J=-1/2}$ Rydberg state by an ultraviolet laser at $\unit[286]{nm}$ with wave vector $\vectorbold{k}_{UV}$, Rabi frequency $\Omega$ and detuning $\Delta$ (see Fig.~1). 
The circular polarized Rydberg laser beam with a waist of $\unit[20]{\mu m}$ is propagating in x-direction, parallel to the magnetic field of $\unit[10]{G}$. 
The ground state atoms are individually trapped in holographically generated two-dimensional optical tweezer arrangements using laser light at $\unit[1064]{nm}$.
In each experimental run about half of the tweezers are randomly loaded with a single atom and the occupation of the traps is detected using fluorescence imaging.
A typical image is shown in Fig.~1c for a 3$\times$16 array. 
In this work we vary the distance of the tweezers between $a=\unit[5]{\mu m}$ and $a=\unit[40]{\mu m}$ (for details of the setup see~\cite{lorenz2021}). 
When the ground state atoms are excited to the Rydberg state, they receive a recoil kick $\vectorbold{p}_r = \hbar \vectorbold{k}_{UV}$, which together with the repulsive ponderomotive force due to the tweezers  leads to efficient ejection out of the trap. 
We detect the lost atoms by comparison of two images, one before and one after the Rydberg laser illumination. 
The duration of the laser pulse is orders of magnitudes shorter than the vacuum limited trap lifetime of about $\unit[80]{s}$, such that lost atoms can be identified with Rydberg excitations.\\

In the limit of a low excitation fraction, any interaction induced line broadening can be understood in a two-atom picture, in which the presence of a contaminating atom leads to a distance-dependent level shift for nearby atoms. 
The normally off resonant laser becomes resonant to the shifted atomic line if the interaction energy matches the detuning and the second atom is excited and subsequently lost from the trap.
The line shifts may be rooted in van-der-Waals or dipolar interactions between two Rydberg atoms or in the electrostatic interaction between a Rydberg atom and an ion~\cite{weller2016, bounds2019}.
In all cases the process can lead to complex kinetically constrained dynamics~\cite{ates2007, garttner2013, lesanovsky2014, mattioli2015,
marcuzzi2016, letscher2017a, klocke2019}, of which signatures have been observed by monitoring the bulk excitation dynamics of a Rydberg coupled gas~\cite{carr2013,  schempp2014, malossi2014, urvoy2015, valado2016, demelo2016, gutierrez2017, letscher2017, helmrich2020, ding2020}.
The incoherent excitation is in contrast to a coherent two-photon excitation of the interacting pair, which becomes resonant at half the detuning.
Facilitated excitation processes reduce the trap lifetime compared to the single atom expectation and imprint characteristic two-body correlations (see Fig.~1d,~e) by which the underlying mechanism can be identified.\\

\begin{figure}[t!!!] \includegraphics[width=\columnwidth]{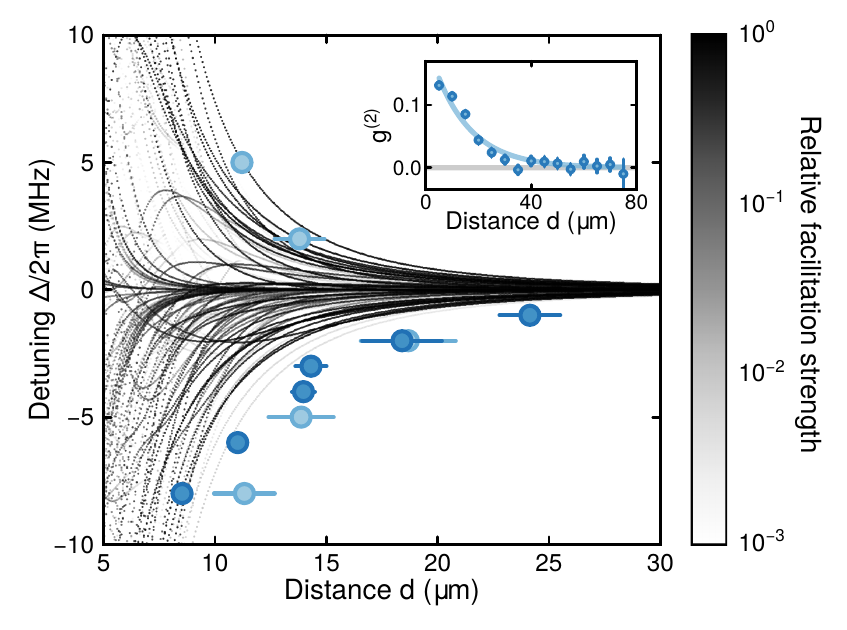}
\caption{ \label{fig:fig2}
Dipolar facilitation range. The figure shows the typical correlation distance
$d_c(\Delta)$ extracted from an exponential fit to $g^{(2)}(d, \Delta)$ for a single line of atoms
(light blue) and a 3$\times$16 array (dark blue), both with spacing
$a=\unit[5]{\mu m}$. Gray shaded lines are the asymptotically
collapsed dipolar pair-potentials with the shading proportional to the expected
logarithmic facilitation strength. 
The inset shows an example of the detected two-body correlations with the exponential fit $g^{(2)}(d, \Delta) \propto \exp(-d/d_c(\Delta))$ to extract $d_c(\Delta)$ for $\Delta = -2 \pi \cdot \unit[5]{MHz}$. 
Error bars denote 1 s.e.m. in the inset and the fit errors in the main panel.
}
\end{figure}

We measure the range of the induced two-point
correlations on lost atoms $g^{(2)}(\mathbf{d}) = \langle (n_{\mathbf{r}} - \langle n_{\mathbf{r}} \rangle) (n_{\mathbf{r} + \mathbf{d}} - \langle n_{\mathbf{r} + \mathbf{d}} \rangle) \rangle$ versus detuning from the atomic resonance. 
Here, $\mathbf{d} = (d_x, d_y)$ is the distance vector connecting the two tweezer positions, $n_{\mathbf{r}}=\pm1$ encodes the occupation of the tweezer at position $\mathbf{r}$ and the averaging is over experimental runs and positions.
We adjusted the Rabi
frequency according to $\Omega/\Delta =  \Omega_m / \Delta_m$ with maximum Rabi
frequency $\Omega_m = 2\pi \cdot \unit[0.4]{MHz}$ at maximum detuning of
$\Delta_m = - 2\pi \cdot \unit[8]{MHz}$, to limit the Rydberg state population
$p_{0} = \alpha(\Delta) \Omega^2/4 \Delta^2$, where $\alpha(\Delta)$ accounts
for excess phase noise of the laser~\cite{si}. We confirmed, that the observed
length scales are constant when decreasing $\Omega$ further. While the
amplitude of $g^{(2)}$ is strongly dependent on the illumination time $\tau$,
we found its spatial dependence to be insensitive to it. In order to assure
comparability of the correlation amplitudes between different settings, we
chose $\tau$ such that 60\% of the initially loaded atoms remained in the
array.\\

For correlations caused by black-body radiation induced state contamination, the length scale of the
correlations $g^{(2)}(d)$ is expected to be set by the pairwise dipolar
interaction potentials. To extract a typical correlation length scale
$d_c(\Delta)$ we fit the data exponentially with $g^{(2)}(d, \Delta) \propto
\exp(-d/d_c(\Delta))$. This empirical fit matches the data well (cf. inset of
Fig.~2). The dipolar interaction potentials are approximately symmetric around the
single atom resonance. In Fig.~2 we show that $d_c(\Delta)$ reproduces this
approximate symmetry and we show that $d_c(\Delta)$ matches with the range of
the dipolar pair-potentials. To illustrate this, we plot the relevant dipolar
potentials of pair-eigenstates $\{\ket{\Psi_2}\}$ asymptotically correlating to
$\ket{r_0, \bar{r}_i}$, with $\ket{\bar{r}_i}$, the i-th state of opposite
parity populated by black-body radiation with rate $\Gamma_{0i}$. The energy
difference $\hbar \Delta_{0i}$ is provided by the black-body photon and, hence,
we plot all pair-states asymptotically at the same energy as $\ket{r_0, r_0}$.
The pair-potentials are shaded according to the relative predicted facilitation
strength. This is defined as the product $\braket{\Psi_2}{r_0,
\bar{r}_i} \Gamma_{0i}/\Gamma_{0k}^\mathrm{max}$ of the overlap of the pair state $\ket{\Psi_2}$
with $\ket{r_0, \bar{r}_i}$ and a normalized black-body coupling rate
$\Gamma_{0i}/\Gamma_{0k}^\mathrm{max}$. The normalization is with respect
to the strongest coupled state with rate $\Gamma_{0k}^\mathrm{max}$. For the
calculations the Pairinteraction~\cite{weber2017} and ARC~\cite{sibalic2017}
software packages have been used. \\

\begin{figure}[t!!!] \includegraphics[width=\columnwidth]{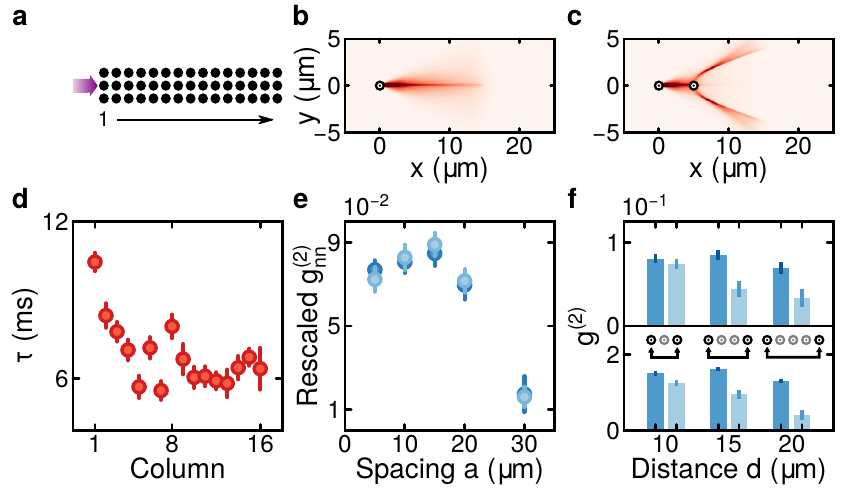}
\caption{\label{fig:fig3}
The role of motion in the facilitation process. \textbf{a.} 
Illustration of the geometry for the 3$\times$16 array including the column numbering 
convention relative to the UV direction (purple arrow).
\textbf{b.} Ensemble of trajectories obtained from classical Monte-Carlo simulations for the motion of the Rydberg atom. 
Shown are all trajectories in a window $\unit[\pm 3]{\mu m}$ around the atomic plane. 
More trajectories crossing a point in space are indicated by lower lightness and the start position of the atoms is marked by the circular symbol. 
The apparent lines in the plot are artefacts of the simulation.
\textbf{c.} Same as in b, but for an additional ponderomotive potential barrier present as indicated by the circular symbol.
\textbf{d.} Column-resolved trap lifetime measurement in a $\unit[5]{\mu
m}$ spaced 3$\times$16 array for $\Omega = 2 \pi \cdot \unit[440]{kHz}$ and
$\Delta = - 2\pi \cdot\unit[4]{MHz}$ . 
\textbf{e.} Next-neighbor correlation $g^{(2)}_{nn}$ in a 1D chain of atoms
versus the array spacing $a$ for detuning $\Delta = -2 \pi \cdot \unit[5]{MHz}$
(light blue) and $\Delta = -2 \pi \cdot \unit[2]{MHz}$ (dark blue).
The data it scaled globally to the same average amplitude.
\textbf{f.} Histograms for $\Delta = -2 \pi \cdot \unit[2]{MHz}$ (top) and $\Delta = -2 \pi \cdot \unit[5]{MHz}$ (bottom) showing the effect of ponderomotive barriers in between two tweezers (see illustration).
The dark (light) blue bars correspond to the case of no (at least one) ponderomotive barrier in between the tweezers at the respective distance.
The correlation is calculated on a subset of the data, in which the tweezers in between (aka the ponderomotive barriers) where empty.
Error bars denote 1 s.e.m.
}
\end{figure}

While our analysis so far confirms that the length scale of the observed
correlations is set by the dipolar pair-potentials, our experiment fails to
reproduce the short distance behavior expected in a picture of fixed atomic
positions (see~\cite{si}). This discrepancy is resolved when considering moving
Rydberg atoms with trajectories determined by the interplay of temperature,
atomic recoil and the tweezer's ponderomotive potential. The atomic recoil
velocity $v_r = \hbar k_{UV}/m = \unit[36]{\mu m/m s}$ for potassium atoms of
mass $m$, due to scattering of photons from the UV laser, is comparable to the typical velocity gained from the ponderomotive potential $v_U = \sqrt{(2 \hbar U / m)} = \unit[40]{\mu m / ms}$, while the thermal velocity of $v_T = \sqrt{k_B T / m} = \unit[6.5]{\mu m/ms}$ for $T = \unit[200]{nK}$ is much smaller.
This results in a directed motion of the atoms exited to Rydberg states (Fig.~3b). The
decay to low lying states takes several hundred microseconds, in which the
Rydberg atoms move by tens of micrometers, clearly invalidating a static
picture~\cite{si}. The impact of this motion can be seen most directly when
analysing the trap lifetime locally (Fig.~3d). Atoms in the first column of the
array (counted with respect to the UV propagation direction) stay almost twice as long
in the trap than atoms in the last column of the array. We attribute this to a
lower effective facilitation rate as no facilitating atoms can approach from
one direction. We confirmed, that the signal is absent without UV illumination. 
To probe for the effect of this motion on the nearest-neighbor
correlations we prepare arrays of different spacing $a$ and compare the
strength of $g^{(2)}_{nn}(a)$ for two different detunings $\Delta =
-2\pi\cdot\unit[2]{MHz}$ and $\Delta = -2\pi\cdot\unit[5]{MHz}$ (see Fig.~3e).
For those detunings, the dipolar potential range differs by almost a factor of
two, but the observed distance dependence of $g^{(2)}_{nn}(a)$ is
indistinguishable. This demonstrates that the sampling of all positions in
flight is hiding the dependence of the dipolar pair-potentials entirely. The
correlations feature a plateau for short distances, which we attribute to the
typical flight distance of the Rydberg atoms within their electronic
lifetime~\cite{si}.\\ 

This result seems to be in contradiction to our observations reported in
Fig.~2, which is resolved when taking into account the presence of other
tweezers in the system. The recoil energy is about $\unit[3]{\mu K}$, comparable to the ponderomotive potential height of $\unit[3.7]{\mu K}$ of the individual tweezers, resulting in a ``shielding'' effect for the
next-nearest-neighbors. To support this interpretation we performed classical simulations, which clearly confirm this effect (see Fig.~3b,c). In Fig.~3f we
show the strength of two-point correlations $g^{(2)}(d)$ for three distances $d$. 
For each distance, we compare the correlation amplitude of nearest-neighbor setting (zero potential barriers in between) to a setting of one or more potential barriers in between.
In all cases, the setting without barriers in between shows strong and almost distance independent correlations.
In contrast, when at least one potential barrier is present, the correlation amplitude decreases with distance and the effect is stronger for the larger detuning, for which the dipolar range is smaller. 
Note that in a static picture no dependence on the presence of empty traps in between the two positions is expected.\\

So far we have focused on the low excitation fraction regime, in which
avalanche facilitation is small. The avalanche effect arises, when an already
facilitated atom is transferred to a state $\ket{\bar{r}_i}$ by black-body
radiation and itself facilitates the excitation of further atoms. To test for
the avalanche mechanism we measure the two-point correlations in the
3$\times$16 geometry with $a=\unit[5]{\mu m}$ for increasing Rabi frequency
while fixing the total fraction of lost atoms. Fig.~4a shows that both, range
and amplitude of two-point correlations increase with higher Rabi frequency.
For the strongest driving, the amplitude of correlation between the two ends
of the array (almost $\unit[80]{\mu m}$ apart) is of comparable strength to
the nearest-neighbor correlation for the weakest drive. The local
measurements reveal, that the strong increase of the $g^{(2)}(d)$ signal is
accompanied by an emergence and subsequent increase of higher order
correlations. In Fig.~4b we show the connected $k$-point correlator at
shortest possible distance ($k$ subsequent tweezers along the UV beam)
$g^{(k)}_{n..n} = \langle \prod_{j=0}^{k} (n_{x + j a} - \langle n_{x + j a}
\rangle) \rangle$. Remarkably, all higher order correlators increase
simultaneously underlining the avalanche character of the process. This is
further supported by a strong broadening of the distribution of lost
atoms~\cite{si}, which is a precursor of the observed bimodality in higher
density settings~\cite{zeiher2016, thaicharoen2018}.

\begin{figure}[t!!!]
\includegraphics[width=\columnwidth]{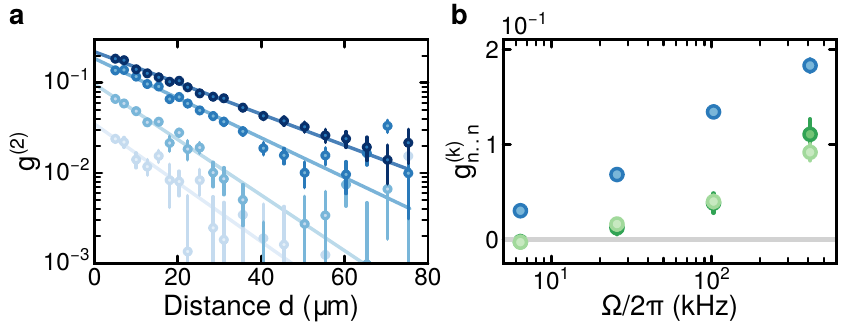}
\caption{\label{fig:fig4} 
Avalanche facilitation. \textbf{a.} Two-point correlations $g^{(2)}$ vs.
distance for different coupling strength in a $\unit[5]{\mu m}$-spaced
3$\times$16 array. The Rabi frequency increases from light to dark blue as
 $\Omega = 2\pi \cdot \unit[(6,25,102,410)]{kHz}$ and the pulse time was adjusted
to fix the fraction of lost atoms to 60\%. Exponential fits are shown as solid lines.
\textbf{b.} Multi-point correlations
vs. coupling strength. The shortest distance connected multi point correlations
$g^{(k)}_{n..n}$ in x-direction are shown for $k=2$ (blue), $k=3$ (light green)
and $k=4$ (dark green). Error bars, where larger than the point size, denote 1
s.e.m.
}
\end{figure}

In this work we microscopically explored correlations emerging in atomic
samples near-resonantly coupled to Rydberg states. We identified
dipolar interactions due to black-body radiation induced state contamination as
the underlying process. Furthermore, we have shown that the recoil triggered,
directed motion of the Rydberg atoms is governing the facilitation process at
nearest-neighbor distance. By increasing the driving strength, our observations
seamlessly connect to previous studies, which concentrated on the avalanche
regime in bulk systems~\cite{goldschmidt2016, desalvo2016, aman2016, gaul2016,
boulier2017, dehond2020}. The possibility to control the Rydberg motion by
repulsive trapping potentials suggests a new possibility to circumvent
catastrophic avalanche dephasing in one- or two-dimensional Rydberg-dressed
systems: With realistic experimental parameters a light-sheet potential at wavelengths
chosen to trap the ground state but to repel the contaminant atoms can be implemented. 
This method works also in combination with tailored trapping wavelengths, which 
allow one to trap the ground state and only one particular Rydberg 
state~\cite{saffman2005a, li2013}. It is compatible with future two-dimensional 
Rydberg quantum processors and simulators and, in particular, it
paves the way to utilize Rydberg dressing for
the design of atomic Hamiltonians for the study of various quantum spin
models~\cite{vanbijnen2015, glaetzle2015,potirniche2017} or to generate useful
states for quantum metrology~\cite{gil2014, kaubruegger2019, kaubruegger2021}.
\\

\begin{acknowledgments}
We acknowledge discussions with S. Hollerith and I. Lesanovsky. This project
has received funding from the European Research Council (ERC) under grant
agreement 678580 (RyD-QMB) and the European Union’s Horizon 2020 research and
innovation program under grant agreement 817482 (PASQuanS). We also acknowledge
funding from Deutsche Forschungsgemeinschaft within SPP 1929 (GiRyd), the MPG
and support from the Alfried Krupp von Bohlen und Halbach foundation.
\end{acknowledgments}

\section*{Supplemental Information}

\renewcommand{\thefigure}{S\arabic{figure}}
\setcounter{figure}{0}

\subsection*{Experimental setting}

We load the potassium-39 atoms from an optical molasses into the tweezer array and subsequently cool them near the motional ground state using Raman sideband cooling. We then ramp the optical tweezers down to 0.5\% of their initial power (corresponding to a trap depth of $\unit[3.7]{\mu K}$). This reduces the inhomogeneities between different tweezers to less than $\unit[50]{kHz}$. Simultaneously, the atoms are adiabatically ``cooled'' to $\unit[200]{nK}$, reducing the Doppler broadening to $2\pi \cdot \unit[50]{kHz} $. This preparation of the atomic sample in the tweezer array is further detailed in reference~\cite{lorenz2021}. 

\begin{figure}[t]
\includegraphics[width=\columnwidth]{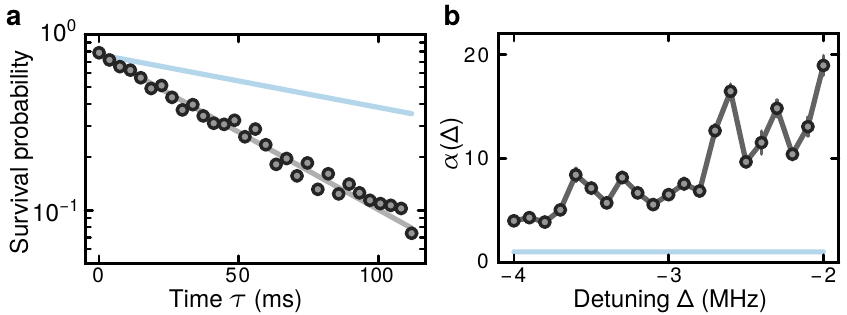}
\caption{\label{fig:fig1} Measurement of the laser noise reduced single atom lifetime. \textbf{a.} Lifetime measurement (gray points) for single atoms with $\unit[30]{\mu m}$ spacing for $\Omega = 2\pi \cdot \unit[0.27]{MHz}$ and $\Delta = -2 \pi \cdot \unit[4]{MHz}$. The gray line is an exponential fit to the data, the blue line is the expected decay for a noise-free laser. \textbf{b.} Reduction $\alpha(\Delta)$ of the single atom lifetime as a function of the detuning. The blue line marks the ideal value $\alpha (\Delta)=1$.}
\end{figure}

\subsection*{Single-atom trap lifetime and laser phase noise}

Spectral components of the laser phase noise that match the detuning resonantly increase the Rydberg population.
The phase noise contribution often dominates the population, in particular, near the resonance.
The impact of the phase noise alone can be conveniently revealed by measuring the single-atom trap lifetime, because any Rydberg excitation is efficiently ejected from the tweezer. 
In the main text we characterize this phase noise by an enhancement $\alpha(\Delta)$ of the Rydberg population w.r.t. the noise free value.
The enhancement factor $\alpha(\Delta) =\tau_{id} / \tau$ follows from the ratio of the measured trap lifetime $\tau$ and the ideal trap lifetime $\tau_{id} = \tau_r \cdot 4 \Delta^2 / \Omega^2$ for a noise-free laser. 
The latter is only limited by the electronic lifetime $\tau_r$ of the Rydberg state.
Here we assumed $\Delta \gg \Omega$.
In the tweezer array, isolated atoms can be realized by placing them far away from each other. 
We use distances of $\unit[30]{\mu m}$ and $\unit[40]{\mu m}$ and confirm that interactions can be neglected in this setting by checking for the absence of correlations in the losses. 

Figure~S1a shows the result of a measurement of the trap lifetime for a Rabi frequency of $\Omega=2\pi
\cdot \unit[266]{kHz}$ and a detuning of $\Delta= - 2\pi \cdot \unit[4]{MHz}$. 
From the exponential fit we extract a lifetime of $\tau = \unit[49.3]{ms} \pm  \unit[1.1]{ms}$, resulting in a factor $\alpha = 2.9$. 
The single-atom prediction shown in figure~1d of the main text includes this factor.
In figure~S1b we show $\alpha(\Delta)$, summarizing the results of all our noise characterization measurements.

\begin{figure}[t]
\includegraphics[width=\columnwidth]{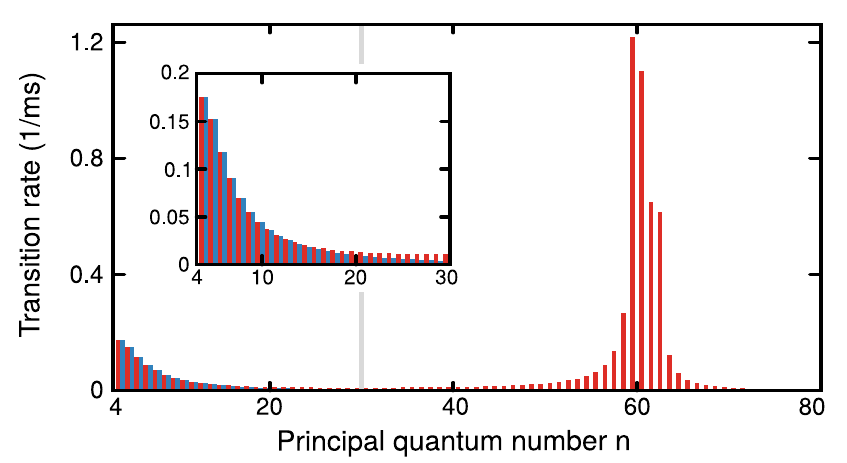}
\caption{
Principal quantum number resolved decay rates of the $62P_{1/2}$ state. The red histogram shows the decay rates to all final states taking into account the black-body radiation background for a temperature of $\unit[300]{K}$. Comparison with the zero temperature histogram (blue) highlights the black-body radiation triggered state changes (see also the inset). These are concentrated around the initial state, where dipole matrix elements are strongest. The gray line marks $n=30$, which we take as a boundary to estimate the total decay rate to facilitating and non-facilitating states (see text). Black-body radiation strongly increases the overall cumulative rate of decay from the initial state.  
}
\end{figure}

\subsection*{Electronic decay of the Rydberg state}

 Figure~S2 shows the decay rates of the chosen Rydberg state $62P_{1/2}$ into all states of different principal quantum number for $\unit[0]{K}$ and for $\unit[300]{K}$. 
 These numbers have been calculated with the ARC software package~\cite{sibalic2017}. 
 At $\unit[300]{K}$ the majority of decays are to nearby Rydberg states. 
 When defining $n=30$ as a boundary between low and high lying states, of which the latter emerge only due to black-body radiation, the ratio of the decay rates is $\sum_{n < 30}{\Gamma_{0,n}} / \sum_{n \geq 30}{\Gamma_{0,n}} = \frac{2\pi \cdot \unit[0.16]{kHz}}{2\pi \cdot \unit[0.8]{kHz}} \approx \frac{1}{5}.$
 
 Atoms, which make a black-body radiation induced transition from the $62P_{1/2}$ to a state of s- or d-orbital symmetry interact via dipolar interactions with atoms in the  $62P_{1/2}$ state. This shifts the transition frequency and is the mechanism behind the facilitated excitation we observe. For the effects of the moving Rydberg atoms and for the avalanche processes at higher driving strength, the time the atoms stay in any Rydberg state is a fundamentally important parameter. This time can be approximated by the zero temperature lifetime, where all decays are to the low lying states. In the vicinity of $n=60$, the $\unit[0]{K}$-lifetime is about $\unit[250]{\mu s}$ for s-states, $\unit[800]{\mu s}$ for p-states and $\unit[500]{\mu s}$ for d-states. Including black-body radiation, none of the states live for more than about $\unit[150]{\mu s}$.  When assuming a lifetime of $\unit[150]{\mu s}$ to roughly estimate the time in which facilitation can take place, the atoms move about $\unit[13]{\mu m}$. Note that this is a crude simplification since the atoms may change their Rydberg state several times before decaying to the ground state.

\subsection*{Dipolar facilitation for fixed atomic positions}

The resonant rate of dipolar facilitation $\gamma_{i}$ due to the i-th pair-potential can be readily calculated in the low excitation fraction regime. It follows from $\gamma_{i} = \Gamma_{\mathrm{eff}} p_0 \Omega^2 \tau_r^2$ with an effective rate $\Gamma_{\mathrm{eff}} = \braket{\Psi_2}{r_0,\bar{r}_i} \Gamma_{0,i}$, taking into account the pair-state overlap $\braket{\Psi_2}{r_0,\bar{r}_i}$ and the black-body coupling rate $\Gamma_{0,i}$. The probability for the atom to be in state $\ket{r_0}$ is given by $p_0 = \alpha(\Delta) \Omega^2/4 \Delta^2$ and includes the laser phase noise factor $\alpha(\Delta)$. The electronic lifetime of the Rydberg state is $\tau_r$.\\

The expected facilitation rate $\gamma_{\mathrm{fac}}$ shown in Figure~S3 for two detunings takes the rates $\gamma_{i}$ of all pair-potentials into account, which become resonant at a certain distance. Additionally, it includes a convolution with a Gaussian of 
standard deviation $\sigma = \unit[0.58]{\mu m}$ to include the thermal extend of the spatial wavefunction in the individual tweezers. The expected spatial dependence is clearly non-exponential in contradiction to our measurements. We interpret this as a further indication for the changing positions of the Rydberg atoms.

\begin{figure}[t]
\includegraphics[width=\columnwidth]{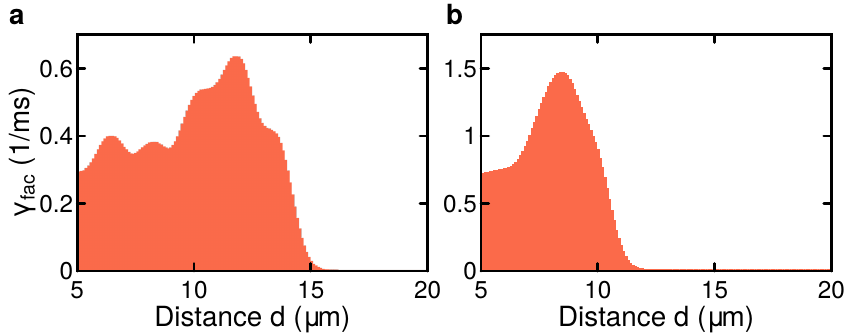}
\caption{
Predicted facilitation rate $\gamma_{\mathrm{fac}}$ assuming fixed positions for the Rydberg atoms for $\Delta = -2\pi \cdot \unit[2]{MHz}$ in \textbf{a} and for $\Delta = -2\pi \cdot \unit[5]{MHz}$ in \textbf{b}. The spatial structure is determined by the crossing of dipolar pair-potentials and the convolution with a gaussian function of standard deviation $\sigma = \unit[0.58]{\mu m}$.
}
\end{figure}

\subsection*{Atom loss distributions at different Rabi frequency and density}

Previous experiments have reported a bimodality in the distribution of lost atoms and interpreted this as a signature of avalanche facilitation triggered by black-body induced state contamination~\cite{zeiher2016, thaicharoen2018}. In figure~S4 we show the distribution of lost atoms for different settings. It increases strongly in width with increasing Rabi frequency and even more when post-selecting the high Rabi frequency data to the high density sector of more than 26 atoms (54\% of the tweezers) loaded. This matches the observation in previous experiments, in which the atomic density was even higher and the avalanche regime was fully realized.  

\begin{figure}[t]
\includegraphics[width=\columnwidth]{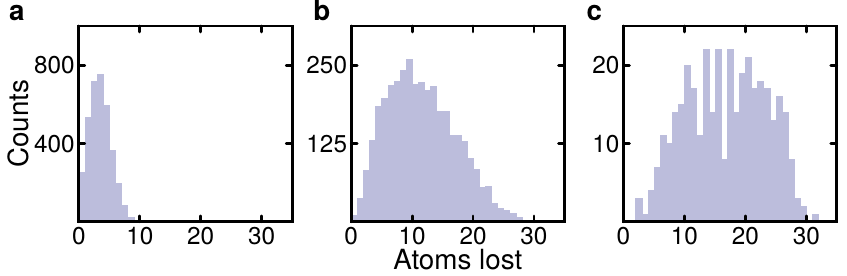}
\caption{
Distribution of lost atoms for strong Rydberg driving at
$\Delta=-2\pi \cdot \unit[3]{MHz}$ and $\Omega=2\pi \cdot \unit[420]{kHz}$.
\textbf{a.} Without UV exposure, where the loss is due to the imaging process.
\textbf{b.}  Exposure time fixed such that 40\% of the atoms are lost. \textbf{c.} Same
settings as for b, but with postselection of the data to more than 26
atoms initially loaded. 
}
\end{figure}

\bibliography{bibliography}

\end{document}